\documentclass[12pt]{article}
\usepackage{amsmath}
\usepackage{amsfonts}
\usepackage{epsfig}
\usepackage{graphicx}
\newcommand{\abar}{\bar\alpha_s}
\newcommand{\tdm}[1]{{\mbox{\boldmath $#1$}}}
\newcommand{\pd}{\partial}

\newcommand{\vk}{\tdm{k}}

\def\phid{\Phi^{\dagger}}
\def\beq{\begin{equation}}
\def\eeq{\end{equation}}

\title{\bf Local one-dimensional reggeon model
of the interaction of several pomerons.}
\author{M.A. Braun, E.M. Kuzminskii, M.I. Vyazovsky\\
Dept. of High Energy physics,
Saint-Petersburg State University,\\
198504 S.Petersburg, Russia}

\begin{document}

\maketitle
{\bf Abstract}

\noindent
We consider the one-dimensional local reggeon theory describing
the leading pomeron with the conformal spin $l=0$
and two subdominant pomerons with $l=\pm 2$.
The dependence of the propagators of pomerons and the $hA$ amplitude
on rapidity are found numerically by integrating the evolution equation.

%%%%%%%%%%%%%%%%%%%%%%%%%%%%%%%%%%%%%%%%%%%%%%%%%%%%%%%%%%%%%%%%%
\section{Introduction}
\noindent
%%%%%%%%%%%%%%%%%%%%%%%%%%%%%%%%%%%%%%%%%%%%%%%%%%%%%%%%%%%%%%%%%%%%
In the framework of the Quantum Chromodynamics in the kinematic region
where energy is much greater than the transferred momenta (''the Regge
kinematics'') the strong interactions can be described by the exchange of
pomerons, which can be interpreted  as bound states of pairs of the
so-called reggeized gluons. In the quasi-classical approximation, which
neglects pomeron loops, this leads to the well-known
Balitsky-Kovchegov equation \cite{bal,kov} and its generalizations
in the form of the JIMWLK equations \cite{jimwlk},
widely used for the description of high energy
scattering. However, it was recognized from the start that at sufficiently high energies
formation of pomeron loops would  change the picture radically.
In the QCD going beyond the quasi-classical approximation and taking account
of loops presents a hardly surmountable problem, which has been not
solved until now. However, this problem could be investigated in simplified models
in which the pomeron is local and interacts with phenomenologically introduced vertices and coupling constants, the Regge-Gribov model
\cite{gribov,migdal,migdal1}. Further simplifying the model to zero transverse dimension ("toy model") one could
study the influence of loops and find that they indeed cardinally change the behaviour at very large energies
\cite{amati1,amati2,jengo,amati3,rossi, braun1}.

It is remarkable that in the QCD the pomeron appears as a set of multiple states, which differ in their energies $\mu$
(intercepts minus unity) and conformal spins. Transformed to the one-dimensional transverse world they appear
as a set of pomerons corresponding to the rising spin $l$ (angular momentum in the full transverse world).
The leading pomeron with $l=0$ phenomenologically may be identified with a local pomeron with  $\mu\simeq 0.12$.
In the QCD language this will correspond to a quite low value of the coupling constant $\bar{\alpha}=0.0433$.
Subdominant pomerons with $l>0$ will  have negative values $\mu$. Of them
the pomeron with $l=\pm 2$ will have the largest $\mu=-0.0531$. It is of some interest to see how the presence of
higher pomerons will influence the behaviour of observables in the limit of high energies.

In this note we generalize the Regge-Gribov model in a zero-dimensional transverse world to include a pair of
subdominant pomerons with $l=\pm 2$. The resulting model includes three types of fields $\Phi_0$, $\Phi_2$ and $\Phi_{-2}$
corresponding to projections of conformal spins $0$ and $2$. It turns out that similar to the model with only one field it admits
numerical study of evolution in energy. As a result we find that loops generated by the additional pomerons act in the same
direction as the main pomeron loops: they diminish both propagators and amplitudes.

%%%%%%%%%%%%%%%%%%%%%%%%%%%%%%%%%%%%%%%%%%%%%%%%%%%%%%%%%%%%%%%%
%%%%%%%%%%%%%%%%%%%%%%%%%%%%%%%%%%%%%%%%%%%%%%%%%%%%%%%%%%%%%%%%%
%%%%%%%%%%%%%%%%%%%%%%%%%%%%%%%%%%%%%%%%%%%%%%%%%%%%%%%%%%%%%
\section{Model}

\noindent
The Hamiltonian of the one-dimensional Regge-Gribov
model with only the pomerons is
\begin{equation}
{H}= - \mu_P \Phi^{+} \Phi
+ i \lambda \Phi^{+}  ( \Phi + \Phi^{+} ) \Phi,
\label{e11}
\end{equation}
where $\Phi,\Phi^{+}$ are the complex pomeron field and its conjugate:
these fields are functions of the rapidity $y$ only.
The mass parameter $\mu_P=\alpha(0)-1$ is defined by the intercept
of the pomeron Regge trajectory and $\lambda$ is the effective coupling
constant.

Our aim is to generalize this model for the case when there are several pomeron states
which interact with a similar triple pomeron interaction. As a guiding model we use the
Balitsky-Kovchegov equation \cite{bal,kov},
which sums fan diagrams with the standard BFKL pomeron and triple pomeron vertices:
\beq
\frac{\partial \Phi(\vk,y)}{\partial y} =
\abar\, (K\otimes \Phi)(\vk,y) -  \abar \Phi^2(\vk,y).
\label{bk}
\eeq
Here the linear terms describe the standard BFKL evolution in the forward direction
\beq
(K\otimes \Phi)(\vk,y) = \frac{1}{\pi}\int \frac{ d^2 \vk'}{(\vk - \vk')^2} \,
\left[
\Phi(\vk',y) - \frac{k^2\Phi(\vk,y)}{\vk'^2 + (\vk - \vk')^2}
\right]
\label{kpom}
\eeq
and $\Phi(\vk,y)=\Phi(-\vk,y)$.

We
develop $\Phi$  in angular momenta $l$, which are actually conformal spins
in  the conformally invariant BFKL dynamics.
\beq
\Phi(\vk,y) = \sum_{l, even}\Phi_l(k,y)e^{il\varphi}.
\eeq
Here $\varphi$ is the angle between $\vk$ and the fixed  direction in the
transverse plane. The conformal spin $l$ goes from $-\infty$ to $+\infty$,
namely $l=...-4,-2,0,2,4...$
(the parity condition requires even  conformal spins).

In absence of interaction evolution of each angular component is given by the solution of the
BFKL equation. For each $l$ the spectrum is continuous and characterized by the conformal parameter $\nu$
with the intercept minus unity given by
\beq
\mu(l,\nu)=\abar\Big[2\psi(1)-\psi\Big(\frac{1+|l|}{2}+i\nu\Big)- \psi\Big(\frac{1+|l|}{2}-i\nu\Big)\Big],\ \ \abar=\frac{N_c\alpha_s}{\pi}.
\label{mu}
\eeq
It goes down with $|l|$ and $\nu$. For $l=0,\pm 2,\pm 4$ at $\nu=0$ one gets
\beq
\mu(0,0)=4\abar\,\ln 2,\ \ \mu(2,0)=4\abar\,(\ln 2-1),\ \
\mu (4,0)=4\abar\, (\ln 2-4/3).
\label{mus}
\eeq
In units $4\abar$ the values are respectively $+0.693$, $-0.307$ and $-0.640$. So only the leading
pomeron with $l=0$ gives  growing amplitudes. Higher pomerons with $|l|\geq 2$ give contributions which diminish with energy,
stronger and stronger as $|l|$ grows. Still one cannot exclude their significant contribution once the interaction is turned on and loops are
taken into account.

We aim to include the first of a series of pomerons with $l=\pm 2$ into a local model in which the dependence on the momenta
are neglected altogether. This is a good approximation for fans in which the momentum is dictated by the nuclear ones, much smaller that the
typical momenta in the BFKL chain. But of course this is a bad approximation for  loops. So the loops we shall study serve only to find and
estimate their influence on fans.  To build a tractable model we also neglect the continuum in the conformal parameter $\nu$ and consider only  $\nu=0$.
Then our pomerons include three discrete states corresponding to $l=0,\pm 2$ which we denote $\Phi_0,\Phi_{\pm 2}$
and the free Hamiltonian is
\beq
H_0=-\mu_0\phid_0\Phi_0-\mu_2(\phid_{-2}\Phi_2+\phid_{2}\Phi_{-2}).
\label{h0}
\eeq
Here we use the notation of the conjugate momentum
\[(\Phi_2)^\dagger =\Phi^\dagger_{-2}, \quad
(\Phi_{-2})^\dagger =\Phi^\dagger_{2},\]
which implies that the subindex $\pm 2$ actually shows the angular momentum transferred by the operator and each term
in (\ref{h0}) is independent of the angle. The energies $-\mu_0$ and $-\mu_2$ are taken from (\ref{mus}).

Now the interaction. From the fan equation (\ref{bk}) we conclude that it has the form
\[\abar\int\frac{d^2k}{4\pi^2}\Big(\Phi^\dagger(\vk)\Phi^2(\vk)+h.c.\Big) , \]
where h.c. means Hermitian conjugation.
For the local reggeons it passes into
\beq
V=\abar\Big(\Phi^\dagger\Phi^2+\Phi{\Phi^\dagger}^2\Big).
\label{v}
\eeq

Taking
\[\Phi=\Phi_0+\Phi_2+\Phi_{-2}\]
and leaving only terms with the total angular momentum zero we get
(suppressing coefficient $\abar$)

\[V=\Big(\Phi_0^\dagger+\Phi^\dagger_2+\Phi^\dagger_{-2}\Big)\Big(\Phi_0+\Phi_2+\Phi_{-2}\Big)^2 +h.c.\]\[
=\Big(\Phi_0^\dagger+\Phi^\dagger_2+\Phi^\dagger_{-2}\Big)\Big(\Phi_0^2+2\Phi_0(\Phi_2+\Phi_{-2})+2\Phi_2\Phi_{-2}\Big)+h.c.\]\beq
=\Phi_0^\dagger(\Phi_0^2+2\Phi_2\Phi_{-2})+2\Phi^\dagger_2\Phi_{-2}\Phi_0+2\Phi^\dagger_{-2}\Phi_2\Phi_0 +h.c.
\label{v1}
\eeq
Together with the standard nonzero commutation relations
\beq
[\Phi_0,\Phi_0^\dagger]=[\Phi_2,\Phi^\dagger_{-2}]=[\Phi_{-2},\Phi^\dagger_2]=1
\label{comrel}
\eeq
the Hamiltonian $H_0+V$ defines a local pomeron model with three pomerons having conformal spins $l=0,\pm 2$.
To preserve the absorptive character of the triple interaction we have to change the real $\abar$ to the imaginary coupling $\abar\to i\lambda$
in correspondence with (\ref{e11}).

\section{The Hamiltonian}
\subsection{Passage to real fields}

\noindent
Our Hamiltonian is the sum of the free Hamiltonian
\beq
H_0=-\mu_0\phid_0\Phi_0-\mu_2(\phid_{-2}\Phi_2+\phid_{2}\Phi_{-2})
\label{h01}
\eeq
and the interaction
\beq
V\!=i\lambda\Big[
\Phi_0^\dagger\Phi_0^2+2\phid_0\Phi_2\Phi_{-2}+2\Phi^\dagger_2\Phi_{-2}\Phi_0+2\Phi^\dagger_{-2}\Phi_2\Phi_0
+{\phid_0}^2\Phi_0+2\phid_2\phid_{-2}\Phi_0+2\phid_0\phid_2\Phi_{-2}+2\phid_0\phid_{-2}\Phi_2\Big] .
\label{vv}
\eeq
To pass to real fields we put
\[
\phid_0=-iu,\ \ \phid_2=-iw,\ \ \phid_{-2}=-iq,\]
\beq
\Phi_0=-iv,\ \ \Phi_{-2}=-ix,\ \ \Phi_2=-it
\label{uwq}
\eeq
with abnormal nonzero commutation relations
\beq
[v,u]=[x,w]=[t,q]=-1 .
\label{crelvzt}
\eeq
In terms of these fields the Hamiltonian becomes real
\beq
H_0=\mu_0 uv+\mu_2(wx+qt),
\label{h0uwq}
\eeq
\beq
V=-\lambda\Big[uv^2+u^2v+2uxt+2wqv+2wxv+2uwx+2qtv+2uqt\Big .]
\label{vuwq}
\eeq

We take the $uwq$ representation in which the state vector is a function of $u,w$ and $q$,
namely $F(u,w,q)$ and the operators $v,x$ and $t$ are the derivatives
\[v=-\frac{\pd}{\pd u},\ \ x=-\frac{\pd}{\pd w},\ \ t=-\frac{\pd}{\pd q} .
\]
Then the Hamiltonian takes the form
\beq
H_0=-\mu_0u\frac{\pd}{\pd u}-\mu_2\Big(w\frac{\pd}{\pd w}+q\frac{\pd}{\pd q}\Big) ,
\label{h02}
\eeq
\beq
V=-\lambda\Big[u\frac{\pd^2}{\pd u^2}-u^2\frac{\pd}{\pd u}+2u\frac{\pd^2}{\pd w\pd q}-2wq\frac{\pd}{\pd u}
+2w\frac{\pd^2}{\pd w\pd u}-2uw\frac{\pd}{\pd w}+2q\frac{\pd^2}{\pd q\pd u}-2uq\frac{\pd}{\pd q}\Big] .
\label{vv1}
\eeq
This form can be used for the power-like or the point-like evolution.

%%%%%%%%%%%%%%%%%%%%%%%%%%%%%%%%%%%%%%%%%%%%%%%%%
\subsection{Transition to two fields}

\noindent
The full Hamiltonian is obviously commutes with the operator of spin
\begin{equation}
l=2 \left( \Phi_{2}^{+}\Phi_{-2} - \Phi_{-2}^{+}\Phi_{2} \right).
\label{e8}
\end{equation}
The conservation of angular momentum allows one to split the solution
into parts with a definite value of $l$, each reducible to only two fields.
Consider a sector with $l=2n\geq 0$. The complete basis of this sector
consists from the Fock basis elements $u^k w^m q^p$ with $m-p=n$.
If we introduce the new zero-spin variable $z=wq$, the basis element
becomes $u^k w^n z^p$ and it is sufficient to consider the action
of derivatives in $w$ and $q$ in the Hamiltonian on it and to express
them via the derivatives in $z$. In particular we evidently find
$$
w\frac{\pd}{\pd w} \left(u^k w^n (wq)^p\right)
=(p+n)\,\left(u^k w^n (wq)^p\right)
= \left(z\frac{\pd}{\pd z} + \frac{l}{2}\right)\left(u^k w^n z^p\right) ,
$$
$$
q\frac{\pd}{\pd q} \left(u^k w^n (wq)^p\right)
=p\,\left(u^k w^n (wq)^p\right)
= z\frac{\pd}{\pd z} \left(u^k w^n z^p\right)
$$
and
\begin{equation}
\frac{\pd^2}{\pd w \pd q} \left(u^k w^n (wq)^p\right)
=(p+n)p\,\left(u^k w^n (wq)^{p-1}\right)
= \left(z\frac{\pd^2}{\pd z^2}
+ \frac{l+2}{2}\frac{\pd}{\pd z}\right)\left(u^k w^n z^p\right) .
\label{e9}
\end{equation}

Using these formulas we can rewrite our Hamiltonian in terms of fields $u$ and $z$.
\beq
H_0=-\mu_0 u\frac{\pd}{\pd u}-2\mu_2 z\frac{\pd}{\pd z}
-\frac{l\mu_2}{2},
\label{h03}
\eeq
\beq
V=-\lambda\Big[u\frac{\pd^2}{\pd u^2}-u^2\frac{\pd}{\pd u}
+l\left(\frac{\pd}{\pd u} - u\right)
+(l+2)u\frac{\pd}{\pd z}
-2z\frac{\pd}{\pd u}+2uz\frac{\pd^2}{\pd z^2}
+4z\frac{\pd^2}{\pd z\pd u}-4uz\frac{\pd}{\pd z}\Big].
\label{vv2}
\eeq
As compared to our original Hamiltonian there appears term $2uz\pd^2/\pd z^2$
corresponding to the quadruple interaction absent in the original formulation.

For $l=-2n<0$ the general basis element is $u^k z^m q^n$
and the analogous consideration shows that the resulting Hamiltonians
are (\ref{h03}) and (\ref{vv2}), where $l$ is substituted with $-l=|l|$.
So the evolution proceeds similarly in the sectors $\pm l$.

One can further simplify the Hamiltonian expressing
\[z=\xi^2.\]
Then we find
\[\frac{\pd}{\pd z}=\frac{1}{2\xi}\,\frac{\pd}{\pd\xi},
\ \ z\frac{\pd}{\pd z}=\frac{1}{2}\xi\frac{\pd}{\pd \xi}\]
and
\begin{equation}
\Big(z\frac{\pd^2}{\pd z^2}\Big)
=\xi^2 \frac{1}{2\xi}\frac{\pd}{\pd \xi}
\Big(\frac{1}{2\xi}\frac{\pd}{\pd\xi}\Big)
=-\frac{1}{4\xi}\frac{\pd}{\pd\xi}+\frac{1}{4}\frac{\pd^2}{\pd\xi^2}.
\label{e10}
\end{equation}
With these formulas we get
\beq
H_0=-\mu_0 u\frac{\pd}{\pd u}-\mu_2\xi\frac{\pd}{\pd \xi}
-\frac{l\mu_2}{2}
\label{h0xi}
\eeq
and
\beq
V=-\lambda\Big[u\frac{\pd^2}{\pd u^2}-u^2\frac{\pd}{\pd u}
+l\left(\frac{\pd}{\pd u} - u\right)
+\frac{(l+1)u}{2\xi}\,\frac{\pd}{\pd \xi}+
\frac{1}{2}u\frac{\pd^2}{\pd \xi^2}-2\xi^2\frac{\pd}{\pd u}
+2\xi\frac{\pd^2}{\pd u \pd\xi}-2u\xi\frac{\pd}{\pd \xi}\Big].
\label{vxi0}
\eeq

Rescaling
\beq
\xi\to\frac{\xi}{\sqrt{2}}
\label{rescale}
\eeq
we finally get (\ref{h0xi}) for $H_0$ and
\beq
V=-\lambda\Big[u\frac{\pd^2}{\pd u^2}-u^2\frac{\pd}{\pd u}
+l\left(\frac{\pd}{\pd u} - u\right)
+\frac{(l+1)u}{\xi}\,\frac{\pd}{\pd \xi}+
u\frac{\pd^2}{\pd \xi^2}-\xi^2\frac{\pd}{\pd u}
+2\xi\frac{\pd^2}{\pd u\pd\xi}-2u\xi\frac{\pd}{\pd \xi}\Big].
\label{vxi}
\eeq
This Hamiltonian is a polynomial in variables and derivatives 
with the total degree not greater than three except for the term
\beq
V_1=-\lambda\frac{(l+1)u}{\xi}\frac{\pd}{\pd\xi} .
\label{v1xi}
\eeq
What is most important it admits the linear transformation
of fields, which eliminates the mixed second derivative.

\subsection{Composite fields}

\noindent
The Hamiltonian (\ref{vxi}) with fields $u$ and $\xi$ in principle
can be used for numerical studies. However, calculations based
on it have a very small region of applicability. Both methods
discussed in Section 4, the power expansion and point-like
evolution, have a rather narrow region of convergence and for the
point-like evolution the blow up starting at already rapidities $8$-$9$.
The reason for this behaviour can be traced to the presence
of the mixed second derivative $\frac{\pd^2}{\pd u\pd\xi}$
in the Hamiltonian (\ref{vxi}). It turns out that by suitably
introducing certain linear combinations of fields $u$ and $\xi$
(''composite operators'') one can exclude this mixed second
derivative and substantially enhance the region of convergence
of our numerical procedure.

We redenote
\[ u\to U,\ \frac{\pd}{\pd u}\to -V,\ \ \xi\to W,\ \ \frac{\pd}{\pd\xi}\to -Z\]
with the anomalous commutators
\[[V,U]=[Z,W]=-1.\]
In terms of these operators
\beq
H_0=\mu_0UV+\mu_2WZ-\frac{l\mu_2}{2}
\label{h0uv}
\eeq
and
\beq
V=-\lambda\Big[-l(U+V)
+UV^2+U^2V-(l+1)\frac{U}{W}Z+
UZ^2+W^2V+2WZV+2UWZ\Big].
\label{vixi}
\eeq

Now we are going to use the composite field operators
to eliminate the mixed second derivative. We define
the new operators as
\beq
u=\frac{U+W}{\sqrt{2}},\ \ v=\frac{V+Z}{\sqrt{2}},\ \
w=\frac{U-W}{\sqrt{2}},\ \ z=\frac{V-Z}{\sqrt{2}}.
\label{uw}
\eeq
With (\ref{uw}) we get
\beq
V=-\sqrt{2}\lambda\Big[uv^2+u^2v+wz^2+w^2z\Big]+V_1+V_2
\label{vinew}
\eeq
with
\beq
V_1=\frac{l+1}{\sqrt{2}}\lambda\frac{u+w}{u-w}(v-z),
\quad
V_2=\frac{l}{\sqrt{2}}\lambda (u+w+v+z) .
\label{v1new}
\eeq
The free Hamiltonian becomes
\beq
H_0=\frac{1}{2}\mu_0(u+w)(v+z)+\frac{1}{2}\mu_2(u-w)(v-z)
-\frac{l\mu_2}{2}.
\label{h0new}
\eeq
Here in the $(uw)$ representation
\[ v=-\frac{\pd}{\pd u},\ \ z=-\frac{\pd}{\pd w}\ .\]
So explicitly the Hamiltonian is now a sum of four terms
\beq
H_0=-\frac{1}{2}\mu_0(u+w)\Big(\frac{\pd}{\pd u}+\frac{\pd}{\pd w}\Big)-\frac{1}{2}\mu_2(u-w) \Big(\frac{\pd}{\pd u}-\frac{\pd}{\pd w}\Big)
-\frac{l\mu_2}{2},
\label{h20fin}
\eeq
\beq
V_I=-\sqrt{2}\lambda\Big[u\frac{\pd^2}{\pd u^2}-u^2\frac{\pd}{\pd u}+w\frac{\pd^2}{\pd w^2}-w^2\frac{\pd}{\pd w}\Big],
\label{v2fin}
\eeq
\beq
V_1=-\frac{l+1}{\sqrt{2}}\lambda\frac{u+w}{u-w}\Big(\frac{\pd}{\pd u}-\frac{\pd}{\pd w}\Big)
\label{v22fin}
\eeq
and
\beq
V_2=\frac{l}{\sqrt{2}}\lambda
\Big[(u+w)-\Big(\frac{\pd}{\pd u}+\frac{\pd}{\pd w}\Big)\Big] .
\label{v221}
\eeq
We expect that in the new form the evolution becomes more stable: the
mixed second derivatives is absent. These expressions in the cases $l=0$
and $l=2$ will be used for the point-like evolution.

%%%%%%%%%%%%%%%%%%%%%%%%%%%%%%%%%%%%%%%%%%%%%%%%%%%%%%%%%%%%%%%%%%%%%%%%%

%%%%%%%%%%%%%%%%%%%%%%%%%%%%%%%%%%%%%%%%%%%%%%%%%%%%%%%%%%%%%%%%%%%%%%%
%%%%%%%%%%%%%%%

\subsection{Power expansion}

\noindent
We present
 \beq
 F(u,w,q;y)=\sum u^iw^jq^k g_{ijk}(y),\ \ i,j,k=0,1,....
 \label{fuwq}
 \eeq
Consider action of different terms in the Hamiltonian $H=H_0+V$
from (\ref{h02}), (\ref{vv1}) on $F$.
We write out only the polynomials in $u,w,q$ omitting the common $g_{ijk}$ and the sign of summation:
 \[-\mu_0u\frac{\pd}{\pd u}=-\mu_0 iu^iw^jq^k,\ \
 -\mu_2\Big(w\frac{\pd}{\pd w}+q\frac{\pd}{\pd q}\Big)=-\mu^2(j+k)u^iw^jq^k,\]
 \[-\lambda u\frac{\pd^2}{\pd u^2}=-\lambda i(i-1)u^{i-1}w^jq^k,\ \
 +\lambda u^2\frac{\pd}{\pd u}=+\lambda iu^{i+1}w^jq^k,\]
\[-2\lambda u\frac{\pd^2}{\pd w\pd q}=-2\lambda jk u^{i+1}w^{j-1}q^{k-1},\ \
+2\lambda wq\frac{\pd}{\pd u}=+2\lambda iu^{i-1}w^{j+1}q^{k=1},\]
\[-2\lambda w\frac{\pd^2}{\pd w\pd u}=-2\lambda iju^{i-1}w^jq^k,\ \
+2\lambda uw\frac{\pd}{\pd w}=+2\lambda j u^{i+1}w^jq^k ,\]
\beq -2\lambda q\frac{\pd^2}{\pd q\pd u}=-2\lambda iku^{i-1}iw^jq^k,\ \
+2\lambda uq\frac{\pd}{\pd q}=+2\lambda k u^{i+1}w^jq^k.
\label{terms}
\eeq

Using these results we present
\beq
HF(u,w,q;y)=\sum u^{i'}w^{j'}q^{k'} f_{i'j'k'}(y) .
\label{hfuwq}
\eeq
Our task is only to relate $i,j,k$ with $i',j'k'$.
For 10 terms in (\ref{terms}) we subsequently write out only these relations and the following
$f_{i'j'k'}$
\[i'=i,\ \ j'=j,\ \ k'=k:\ \ -\mu_0i'g_{i'j'k'},\]
\[i'=i,\ \ j'=j,\ \ k'=k;\ \ -\mu_2(j'+k')g_{i'j'k'},\]
\[i'=i-1,\ \ i=i'+1,\ \ j'=j,\ \ k'=k:\ \  -\lambda i'(i'+1)g_{(i'+1)j'k'},\]
\[i'=i+1,\ \ i=i'-1,\ \ j'=j,\ \ k'=k:\ \ +\lambda(i'-1)g_{(i'-1)j'k'},\]
\[i'=i+1,j'=j-1,\ \ k'=k-1,\ \ i=i'-1,\ \ j=j'+1,\ \ k=k'+1:\]\[ -2\lambda (j'+1)(k'+1)g_{(i'-1)(j'+1)(k'+1)},\]
\[i'=i-1,\ \ j'=j+1,\ \ k'=k+1,\ \ i=i'+1,\ \ j=j'-1,\ \ k=k'-1:\]\[ +2\lambda(i'-1)g_{(i'+1)(j'-1)(k'-1)},\]
\[i'=i-1,\ \ i=i'+1,\ \ j'=j,\ \ k'=k:\ \ -2\lambda(i'+1)j'g_{(i'+1)j'k'},\]
\[i'=i+1,\ \ i=i'-1,\ \ j'=j,\ \ k'=k:\ \ +2\lambda j' g_{(i'-1)j'k'},\]
\[i'=i-1,\ \ i=i'+1,\ \ j'=j,\ \ k'=k:\ \ -2\lambda (i'+1)k'g_{(i'+1)kj'k'},\]
\[i'=i+1,\ \ i=i'+1,\ j'=j,\ \ k'=k:\ \ +2\lambda k'g_{(i'-1)j'k'}.\]

So in the $g_{ijk}$ representation we find
$Hg=f$,
where
\[f_{ijk}=-\mu_0ig_{ijk}-\mu_2(j+k)g_{ijk}
-\lambda i(i+1)g_{(i+1)jk}+\lambda(i-1)g_{(i-1)jk}\]\[
-2\lambda(j+1)(k+1)g_{(i-1)(j+1)(k+1)}+2\lambda(i+1)g_{(i+1)(j-1)(k-1)}\]\beq
-2\lambda(i+1)jg_{(i+1)jk}+2\lambda jg_{(i-1)jk}
-2\lambda(i+1)kg_{(i+1)jk}+2\lambda kg_{(i-1)jk}.
\label{hg}
\eeq
This expression is the basis for the power-like evolution.

\section{Numerical results}

\noindent
Our numerical calculations of the propagators and amplitudes 
consist in solving the evolution equations in rapidity
by the Runge-Kutta method starting from their values at $y=0$ 
\beq
\frac{\pd F(y)}{\pd y}=-HF(y) ,
\label{df}
\eeq
where $F(y)$ is the wave function at rapidity $y$.
In fact it is a function of field variables. For instance,
in variables $u,w$ and $q$ we have the wave function $F(u,w,q;y)$
and the Hamiltonian is given by (\ref{h02}) and (\ref{vv1}).
For the point-like approach we introduce a lattice for all field
variables with $N$ points in each variable and consider the wave
function evolution on this lattice, treating all derivatives
in the Hamiltonian as discretized. For power expansion the evolving
quantities are defined by (\ref{fuwq}) and the corresponding
Hamiltonian determined by (\ref{hg}).

For the propagators the initial values of the wave function are given
by $F=u$, $F=w$ or $F=q$ for pomerons with spins $0$, $+2$ and $-2$,
respectively. The amplitudes are found from the same initial condition
by projecting on the relevant final states as explained
in \cite{braun1,bkv}.

\subsection{Propagators}

\begin{figure}[h!p]
\begin{center}
\includegraphics[width=10 cm]{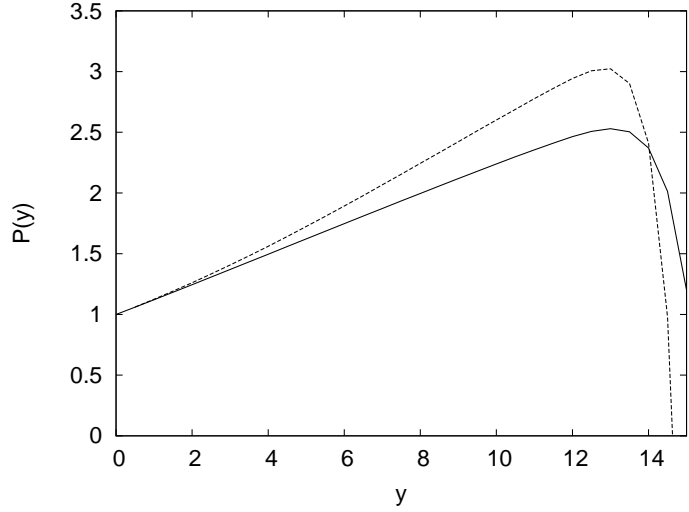}
\caption{The propagator of the pomeron $l=0$ at different rapidities found by power expansion
with three pomerons  $l=0,\pm 2$ (lower curve) and without subdominant pomerons of
$l=\pm 2$ (upper curve). $\mu_0=0.12$, $\mu_2=-0.0531$, $\abar= 0.0433$.}
\label{fig1}
\end{center}
\end{figure}

\noindent
With three fields $\Phi_0$, $\Phi_{+2}$ and $\Phi_{-2}$
numerical evolution both by power expansion and by points turns out to be possible
only for a limited interval of rapidities $0<y<y_{max}$ beyond which evolution breaks down.
In our power expansion we limited the powers of $u$, $w$ and $q$ by the maximal number 20
to perform evolution in a reasonable amount of time. The resulting
maximal rapidity turned out to lie in the region $14$-$15$. Remarkably, this restriction
is the same as with the pure Gribov model without any extra field. This is clearly seen
in Fig. \ref{fig1} where we show both the results found by power expansion with three fields
(the lower curve) and with only the main pomeron $l=0$ (the upper curve). As expected,
the extra loops with pomerons $l=\pm 2$ bring the propagator further down, the drop
growing with rapidity.

\begin{figure}[h!p]
\begin{center}
\includegraphics[width=10 cm]{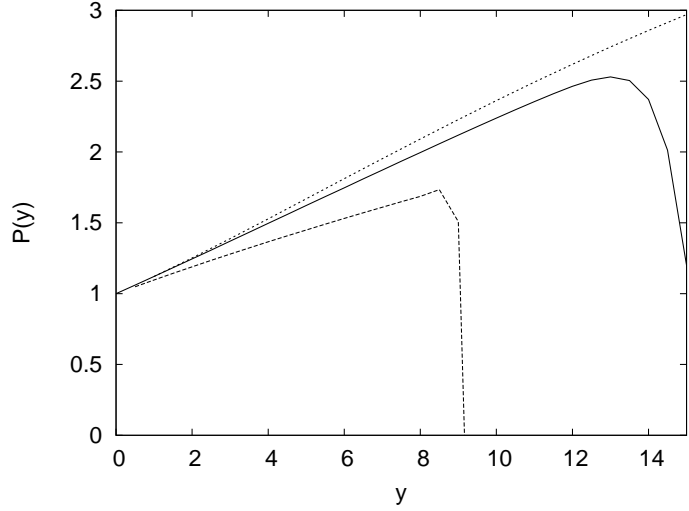}
\caption{The propagator of the pomeron $l=0$ at different rapidities
with three pomerons $l=0,\pm 2$ obtained by power expansion (middle curve),direct point-like evolution (lower curve) and
two-field point-like evolution with composite fields (upper curve).
$\mu_0=0.12$, $\mu_2=-0.0531$, $\abar= 0.0433$.}
\label{fig2}
\end{center}
\end{figure}

The point-like evolution of course gives practically identical results with the Hamiltonians
(\ref{h02}) and (\ref{vv1}) with three fields and (\ref{h03}) and (\ref{vv2}) for $l=0$ with two fields
(using $N=400$ points), although the processor time is naturally much shorter with two felds.
In both cases for the point-like evolution convergence is the same and considerably worse than for the power-like expansion.
The maximal rapidity goes down to $8$-$9$
above which the evolution breaks down. This is a new phenomenon, absent for a single pomeron,
where evolution is possible to very high rapidities. Presumably this effect is due to the presence
of the mixed second derivatives in the Hamiltonians. Numerically the propagator obtained by the point-like evolution
is somewhat smaller (in the convergence interval) than the one obtained by power expansion. This is illustrated
in the bottom curve in Fig. \ref{fig2} to compare with the power-like evolution shown by the middle curve.

\begin{figure}[h!p]
\begin{center}
\includegraphics[width=10 cm]{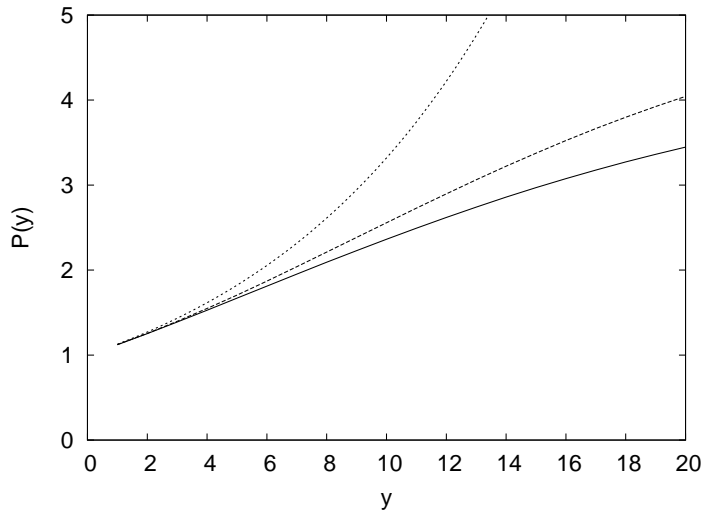}
\caption{The propagator of the pomeron $l=0$ at different rapidities
with three pomerons  $l=0,\pm 2$ obtained by
two-field point-like evolution with composite fields (lower curve) and without the subdominant pomerons (middle curve).
$\mu_0=0.12$, $\mu_2=-0.0531$, $\abar= 0.0433$.
The upper curve corresponds to the free propagator $\exp(0.12 y)$.}
\label{fig3}
\end{center}
\end{figure}

Passage to composite fields and thus elimination of the mixed derivative makes evolution drastically better and possible
in a wide interval of rapidity up to very large rapidities.  The results found for the main pomeron propagator by this methods are shown
in the same Fig. \ref{fig2} by the upper curve and separately in Fig. \ref{fig3} for a wider interval of $y$
together with the results found without the subdominant pomerons.

\begin{figure}[h!p]
\begin{center}
\includegraphics[width=10 cm]{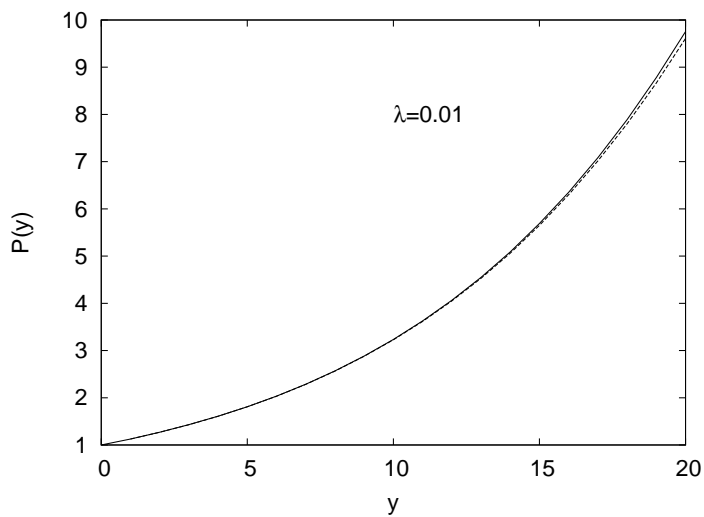}
\caption{The propagator of the pomeron $l=0$ at different rapidities
with three pomerons $l=0,\pm 2$ obtained by power expansion (upper curve) and
two-field point-like evolution with composite fields (lower curve) with a smaller
coupling constant $\abar=0.01$ and the same
$\mu_0=0.12$ and $\mu_2=-0.0531$.}
\label{fig4}
\end{center}
\end{figure}

In relation to Figs. \ref{fig1} and \ref{fig2} we have to note that the precision of the power expansion drops with the growth of $\lambda$.
From our previous experience \cite{bkv} the value
$\lambda=0.0433$ used in our calculations lies quite close to the convergence border $\lambda=0.04$. To illustrate this point  in Fig.\ref{fig4}
we compared the results found by power expansion and point-like evolution with composite fields at considerably smaller value $\abar=0.01$.
One observes a nearly complete agreement.

\begin{figure}[h!p]
\begin{center}
\includegraphics[width=10 cm]{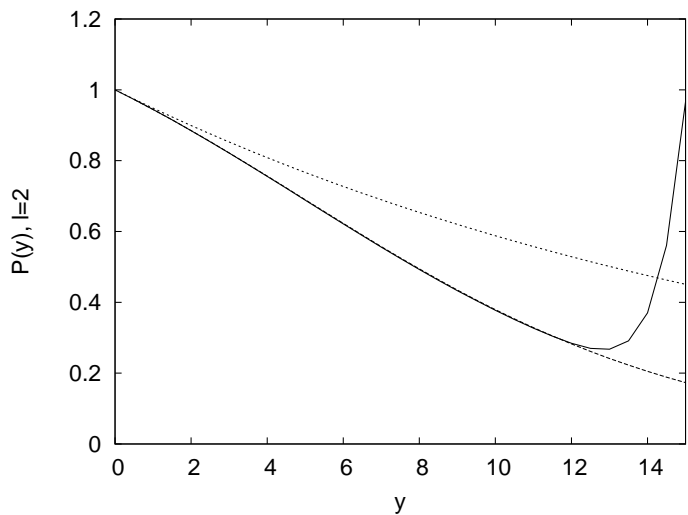}
\caption{The propagator of the subdominant pomeron $l=2$ at different rapidities
with three pomerons $l=0,\pm 2$ obtained by power evolution (middle curve) and
two-field point-like evolution with composite fields (lower curve).
$\mu_0=0.12$, $\mu_2=-0.0531$, $\abar= 0.0433$.
The upper curve corresponds to the free propagator $\exp(-0.0531 y)$.}
\label{fig5}
\end{center}
\end{figure}

For the subdominant pomerons we meet with a certain trouble. For $l=2$ the power expansion works as for the main pomeron, but the point-like evolution with composite fields
converges only when the number of variable points does not exceed $N=260$. For larger $N$ evolution breaks down already at $y=3$.
In the region of convergence the power expansion and point-like evolution with composite field give practically identical results up to $y\sim 12$, starting
from which the power expansion blows up, which apparently shows the breakdown of convergence. This is illustrated in Fig. \ref{fig5}.
Note that loops diminish the propagator with $l=2$ in the same way as with $l=0$, as follows comparing with the upper curve in Fig. \ref{fig5}.

\newpage
\subsection{Amplitude}

\noindent
Our amplitude depends on the initial and final coupling constants for the pomerons.
To couple the pomeron with $l=\pm 2$ to hadrons one should assume that they possess some quadrupole moment.
For spherical hadrons the quadrupole moment is absent  and for any hadrons it should be very small.
So we neglect the coupling of pomerons with $l=\pm 2$ to nucleons and our amplitude will take them into account only in the loops.

To calculate the amplitude we have to know the two coupling constants of the principal pomeron with $l=0$, $g_{ini}$ for the incoming hadron
(proton) and $g_{fin}$ for the final amplitude. The latter corresponds to the coupling to the nucleon inside
the nucleus. The corresponding fan diagram describes the amplitude at fixed impact parameter $b$ and is equal
to ~\cite{schwimmer}
\beq
{\cal A}_A(y,b)=\frac{g_{ini}g_{fin}(A,b)e^{\mu y}}{1+\lambda g_{fin}(A,b)\frac{e^{\mu y}-1}{\mu}}.
\label{schwim}
\eeq
The final coupling constant depends on the atomic number $A$ and impact parameter $b$. For simplicity
we assume that the nucleus is a sphere of radius $R_A=A^{1/3}R_0$. Therefore for collisions with the nucleus
$g_{fin}\propto\theta(R_A-b)$ and does not depend on $b$ as soon as $b<R_A$. As to $A$-dependence
it is natural to assume (see also ~\cite{schwimmer}) that $g_{fin}\propto A^{1/3}$. So in the end
we take
\beq
g_{fin}=A^{1/3}g_{ini}\theta(R_A-b).
\label{gfin}
\eeq
The fan amplitude of $hA$ interaction obtained after integration over $b$ becomes
\beq
{\cal A}_A(y)=\pi A R_0^2\frac{g_{ini}^2 e^{\mu y}}{1+\lambda A^{1/3}g_{ini}\frac{e^{\mu y}-1}{\mu}}.
\label{schwim1}
\eeq

With a small $\lambda$ at comparatively low energies $\mu y\sim 1$ we find for the scattering on the nucleus
\beq
{\cal A}_A(y)=\pi A R_0^2g_{ini}^2 e^{\mu y},\ \ \lambda e^{\mu y}<<1
\label{schwim2}
\eeq
(single pomeron exchange). For the scattering on the proton we guess that one has to put $A=1$ and change the nuclear radius $R_A$
to the  proton radius $R_p$. In this way at comparatively low energies with the proton target one gets
\beq
{\cal A}_{pp}(y)=\pi R_p^2g_{ini}^2 e^{\mu y},\ \ \lambda e^{\mu y}<<1.
\label{schwim3}
\eeq
We can use this formula to fix the value of $g_{ini}$.
At $s=100$ GeV$^2$ the cross-section (which is just ${\cal A}$ in our normalisation) is approximately
equal 40 mbn=4 fm$^2$. At this energy $y_0=\ln(s/s_0)=4.6$ where we lake $s_0=1$ Gev$^2$. We also take $R_p=0.8$ fm.
With these values and taking $\mu=0.12$
\beq
g_{ini}^2=\frac{{\cal A}_{pp}(4.6)e^{-\mu y_0}}{\pi R_p^2}=\frac{4*e^{-0.12*4.6}}{\pi*0.64}=1.144,\ \ g_{ini}=1.069.
\label{gini}
\eeq

\begin{figure}[ht]
\begin{center}
\includegraphics[width=10 cm]{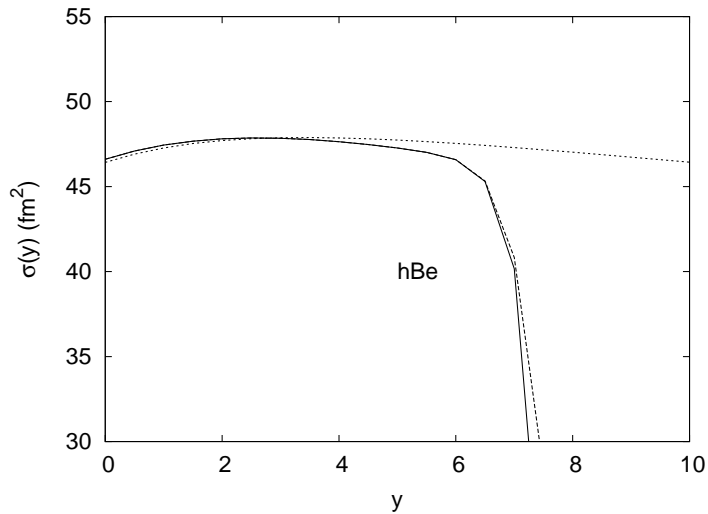}
\caption{The hBe cross-sections calculated by power expansion (the leftmost curve), power expansion of the inverse amplitude
(the curve slightly shifted to the right) and point-like evolution with composite operators (the smooth upper curve).
$\mu_0=0.12$, $\mu_2=-0.0531$, $\abar= 0.0433$.}
\label{fig6}
\end{center}
\end{figure}

From the fan diagrams we can conclude that the effective coupling constant in the interaction with nuclei is actually
given by $\lambda g_{ini}A^{1/3}$ and so grows with $A$. So passing to the diagrams with loops we expect that
in the power expansion convergence will be poorer with the growth of $A$. Calculations show that this is indeed the case.
In Fig. \ref{fig6} we show the cross-sections for the scattering on Berillium (A=9) found by the power expansion and point-like
evolution with composite operators. The two curves corresponding to the power expansion refer to direct expansion and expansion
of the inverse amplitude. In both cases we observe a breakdown of convergence at $y\sim 7.7$ (slightly higher for the expansion
of the inverse amplitude). The situation for heavier nuclear targets shows even worse convergence for the power expansion.

Because of this our results for the $hA$ cross-sections presented in the following figure refer to constructive calculations
by the point-like evolution with composite operators.
We separate dimensionful factor $\sigma_A(y=0)=\pi R_0^2 A$ carrying the bulk of the $A$-dependence and present the ratio
\beq
R(y)=\frac{\sigma_A(y)}{\sigma_A(y=0)}
\eeq
showing the energy-dependence of the cross-sections.
To illustrate the influence of the subdominant pomerons and loops in general
we also present the results obtained with only the dominant pomeron $l=0$ and without loops (fan diagrams). As we see in all cases
the loops diminish the amplitude and the effect of the subdominant pomerons acts in the same direction as of the dominant one.

\begin{figure}[h!p]
\begin{center}
\includegraphics[width=8 cm]{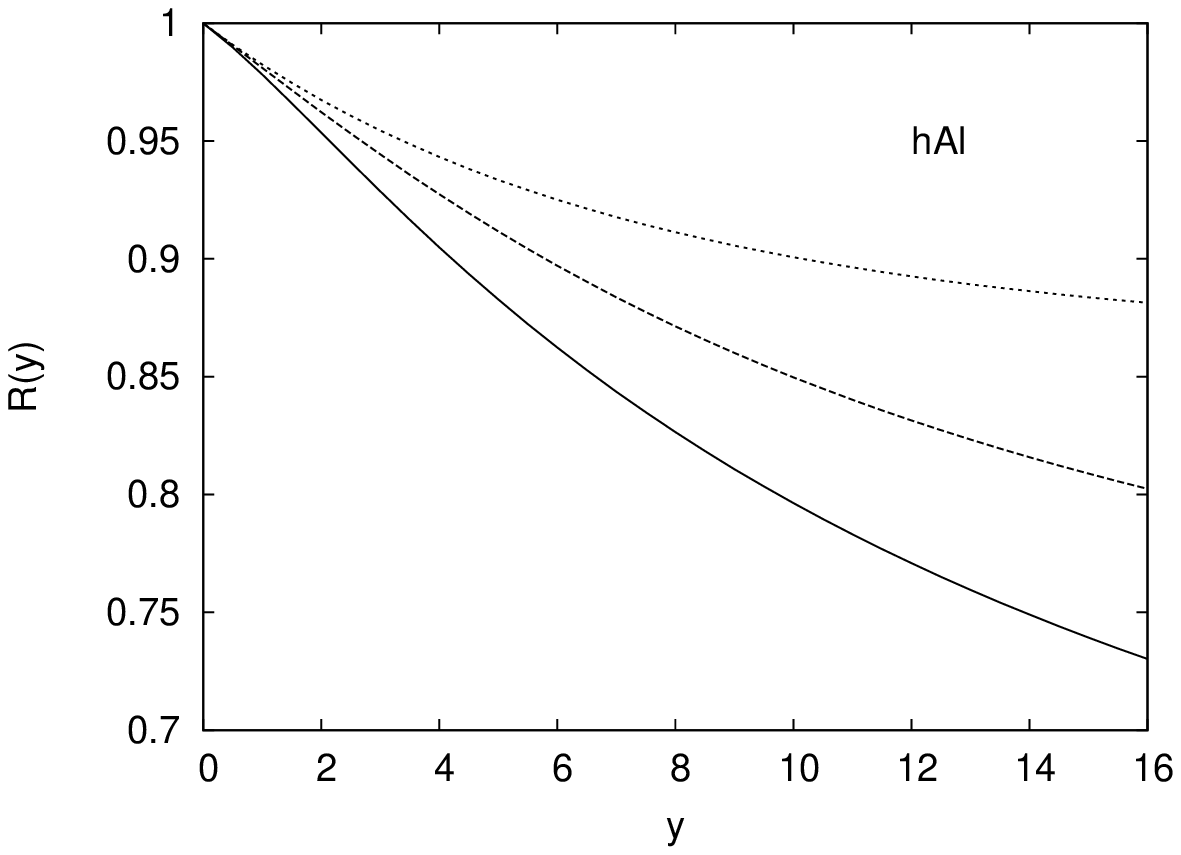}
\includegraphics[width=8 cm]{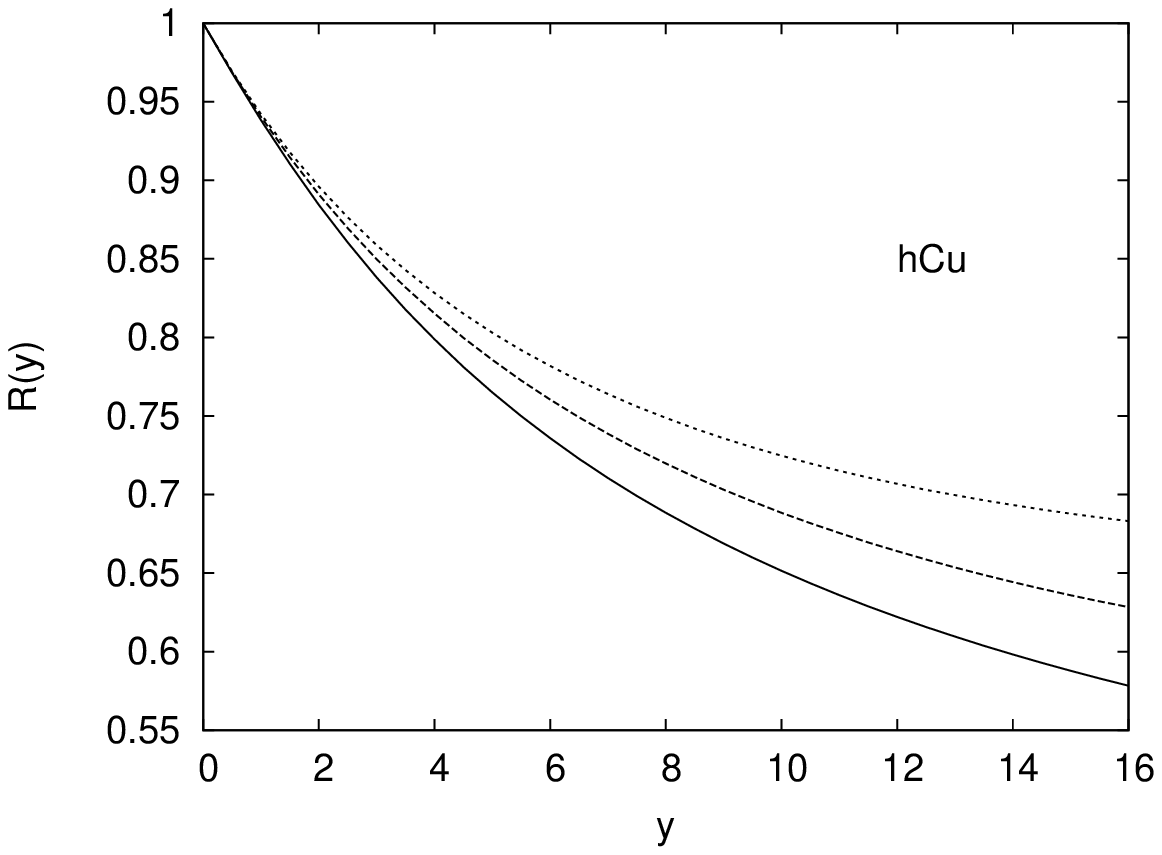}
\includegraphics[width=8 cm]{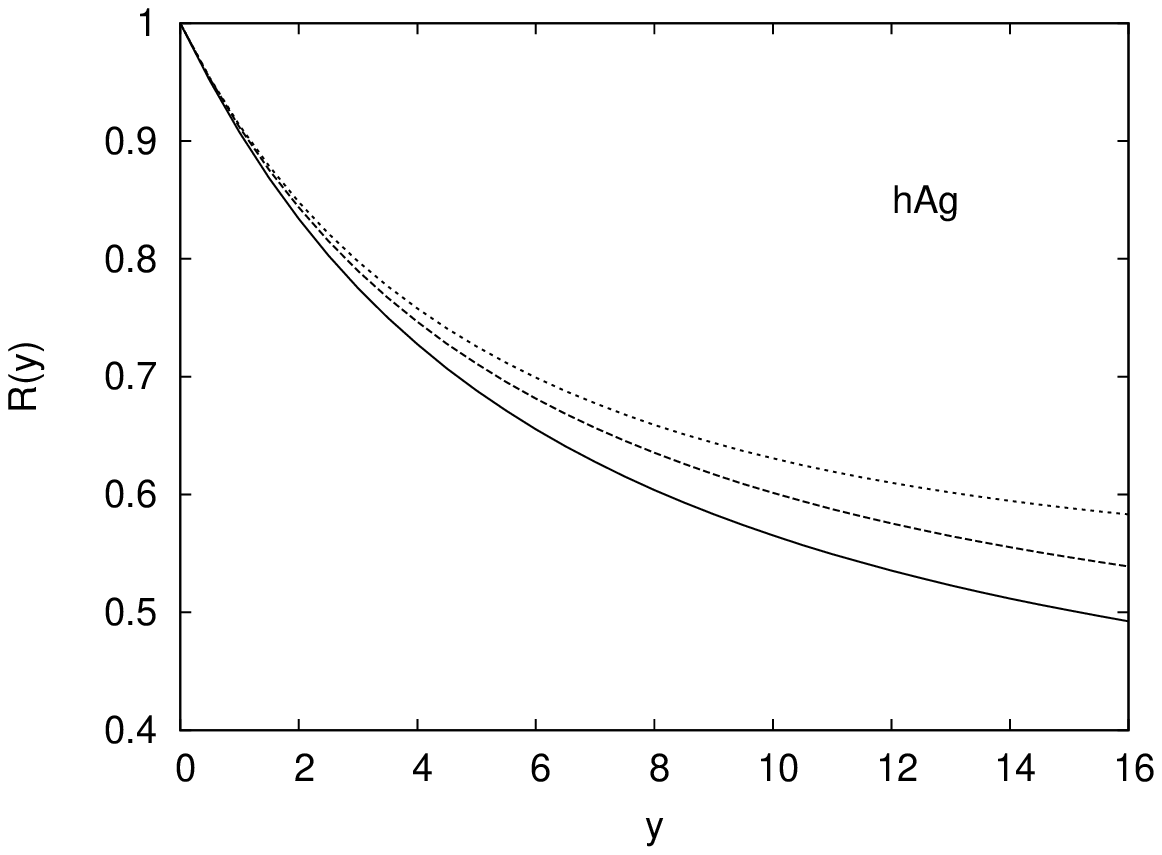}
\includegraphics[width=8 cm]{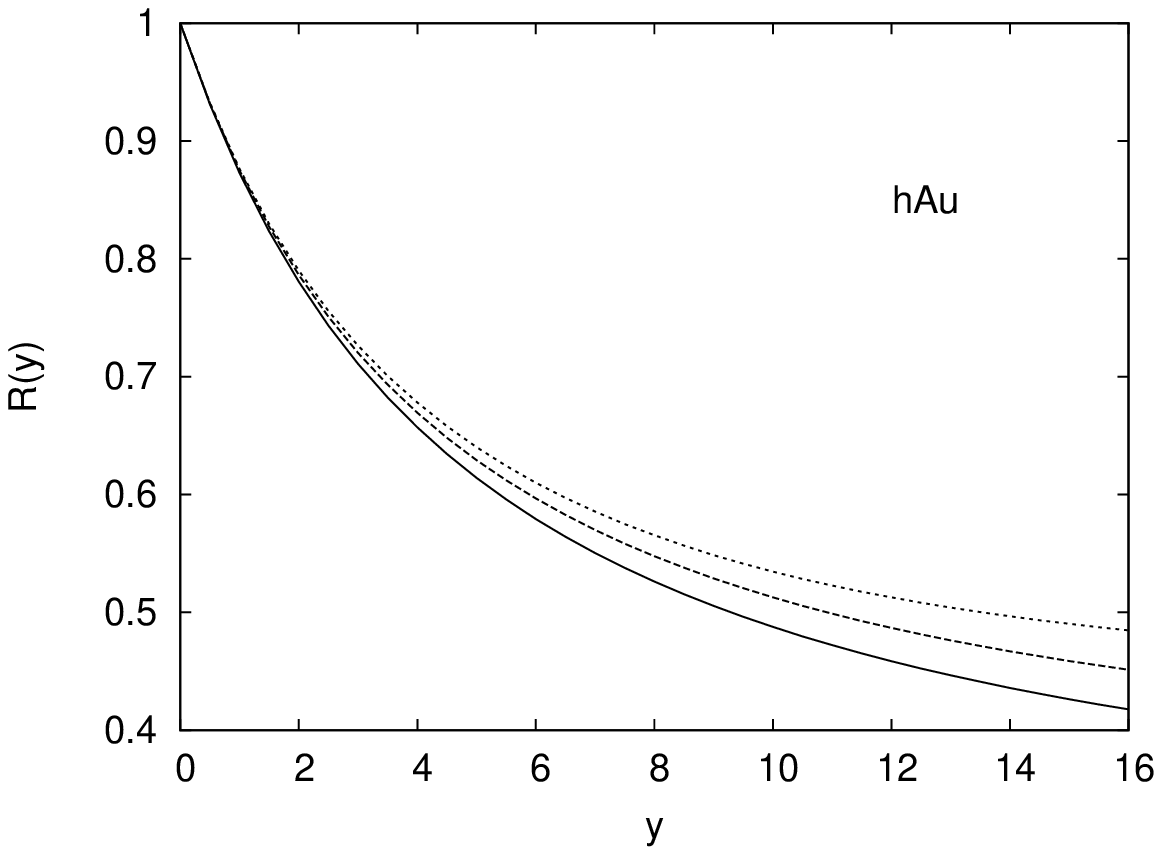}
\caption{The $y$-dependence of the $hA$ cross-sections calculated by point-like evolution with composite operators (the lower curves),
without the subdominant pomerons $l=\pm 2$ (the middle curves) and without loops (fan diagrams, the upper curves).
Panels correspond to the scattering on Aluminium (upper left), Copper (upper right), Silver (bottom left) anf Gold (bottom right).
In all cases
$\mu_0=0.12$, $\mu_2=-0.0531$, $\abar= 0.0433$.}
\label{fig7}
\end{center}
\end{figure}

\newpage
\section{Conclusions}

\noindent
We constructed  a  local pomeron model in the one-dimensional transverse world with
three interacting pomerons of conformal spins 0 and 2 and phenomenological intercepts
and coupling. Evolution in rapidity was studied numerically. It was found that direct evolution from the
Hamiltonian of the three fields is only possible in a restricted interval of rapidities below $y\sim 6-12$
due to divergence. This obstacle was overcome by passage to two effective fields for sectors with $l=0$ and
$l=2$ separately. Transition to composite fields then allows evolution in a wider interval, up to $y=20$
considered in the calculations.

The resulting propagators and $hA$ amplitudes were found to be smaller than without subdominant pomerons with
$l=2$ and, as expected, without any loops at all. So the subdominant loops were found to act in the same direction as the 
dominant ones corresponding to $l=0$. This result shows the difference between the pomeron with signature $+1$ and the odderon with signature $-1$.
The odderon loops in contrary act in the opposite direction and enhance both the propagators and amplitudes ~\cite{bkv}. 
%%%%%%%%%%%%%%%%%%%%%%%%%%%%%%%%%%%%%%%%%%
%%%%%%%%%%%%%%%%%%%%%%%%%%%%%%%%%%%%%%%%

\end{document}